\documentclass[aps,pra,reprint,amsmath,superscriptaddress,onecolumn]{revtex4-2}
\usepackage{graphicx}
\usepackage{dcolumn}
\usepackage{epstopdf}
\usepackage{xcolor}

\begin{document}

\title{Measuring the local mechanical properties of a floating elastic sheet}
%
%
	\author{G. Le Doudic}
\author{M. Jafari}%
\author{J. Barckicke}%
\author{S. Perrard}%
\author{A. Eddi}%
 \email{antonin.eddi@espci.fr}
\affiliation{%
PMMH Lab, ESPCI, CNRS, PSL University, Sorbonne Universit\'e, Universit\'e Paris Cit\'e\\
7 quai Saint Bernard, 75005 Paris, France
}%

	
	\date{\today}

\begin{abstract}
Polar regions are covered by sea ice, which can be seen as a thin solid elastic sheet with heterogeneous mechanical properties. The dynamics of deformation of a floating solid sheet are primarily governed by gravity, water density, and the flexural modulus, which depends on its mechanical properties, namely the thickness, the Young's Modulus and the Poisson ratio. Non-invasive methods from seismology can retrieve these three parameters from sheet deformation dynamics. In this article, we developed another method to extract locally the flexural modulus of a floating thin elastic sheet from the spatio-temporal deformations of the sheet. We perform laboratory experiments to test the accuracy and the robustness of this method on silicon membranes of controlled mechanical properties. Using patches of different thicknesses and shapes, we eventually draw maps of sheet thickness, with a sub-wavelength spatial resolution. 
\end{abstract}
\maketitle
	 
	

\section*{Introduction}

    Sea ice seasonally covers the polar oceans \cite{Cavalieri_2012}, creating a heterogeneous solid crest, much thinner than its horizontal extension. These zones are dynamically active, with many complex physical processes at play, that lead to significant spatial variations of their mechanical properties at all scales. Field measurements of sea ice thickness have been performed, but the spatial variations of the mechanical properties remain mainly unknown. There is a need for noninvasive methods, that can monitor the mechanical properties of ice at the scale of the ice deformations (10 meters and above). Seismic measurements in particular have been introduced~\cite{Marsan2012} to assess the Young's modulus, Poisson ratio and ice thickness. These techniques are based on the measurement of the various elastic waves that propagate within a thin elastic sheet such as sea ice. Waves associated with in-plane motions are non-dispersive at low frequencies, and depend on Young's modulus and Poisson ratio, while waves associated with out-of-plane local displacements are dispersive in nature, and also depend on the plate thickness. From either active noise sources or passive recording of natural seismic activity, measuring the wave dispersion relations provides access to the mechanical properties of ice. Further development, based on fitting the wave shape, also gives access to the averaged plate density~\cite{Moreau2020, Serripierri2022}. These methods, however, measure spatially averaged values of the mechanical parameters. Accessing the local parameter values and their spatial variations remains an ongoing challenge. \\ 
    
    The hydro-elastic waves associated with out-of-plane local displacements have been also studied at the reduced scale of the laboratory, using mimetic materials such as rubber sheets~\cite{Domino_2018}, granular rafts~\cite{Planchette2012} or at the intermediate scale of a wave basin, using PVC disks~\cite{Montiel2013HEW1}.
    These hydro-elastic out-of-plane waves exhibit two regimes: gravity dominates at lower frequencies while bending dominates at higher frequencies.
    Both field measurements and lab experiments exhibit these two regimes.
    At the laboratory scale, non-invasive space- and time-resolved wave measurements have been developed, offering improved resolution of wave propagation. Doing so, classical wave effects such as diopter, lenses~\cite{Domino_2018}, and periodic materials~\cite{Domino_2020} have been observed and quantitatively measured with hydro-elastic waves. The laboratory analogue can then be seen as a platform for developing methods that could potentially be deployed for the analysis of field data. \\

    The presence of spatial variations of mechanical properties is known to affect wave propagation. Such variations are used in optics and acoustics, for instance, to infer the local properties of biological tissues, or in seismology to identify underground veins. A local technique in particular has been developed to measure the local phase velocity of non-dispersive waves, such as acoustic waves ~\cite{Ing2010, Etaix2010}, and shallow water surface waves ~\cite{Przadka2013}. However, for deep water waves and hydro-elastic waves, which exhibit dispersion, there is currently no equivalent technique available.\\

    In the paper, we aim to develop a technique to extract the local values of the mechanical properties of a floating membrane, using spatiotemporal measurements of the wave height. To do so, we perform laboratory scale experiments, using thin elastic sheets that have been previously well characterized~\cite{Domino_2018}. The paper is organized as follows: we first present the experimental setup and the surface reconstruction technique. We demonstrate our ability to quantitatively measure the mechanical properties of a uniform floating elastic sheet. We then present the k-extraction method and its practical implementation. Next, we show that this method extracts the local elastic properties of the sheet in controlled heterogeneous configurations. Last, we show that in the flexural wave regime, we can quantitatively extract local mechanical parameters and we eventually draw thickness maps of heterogeneous sheets.

\section{Experimental set-up}

\begin{figure}[t]
    \centering
\includegraphics[width=.7\columnwidth]{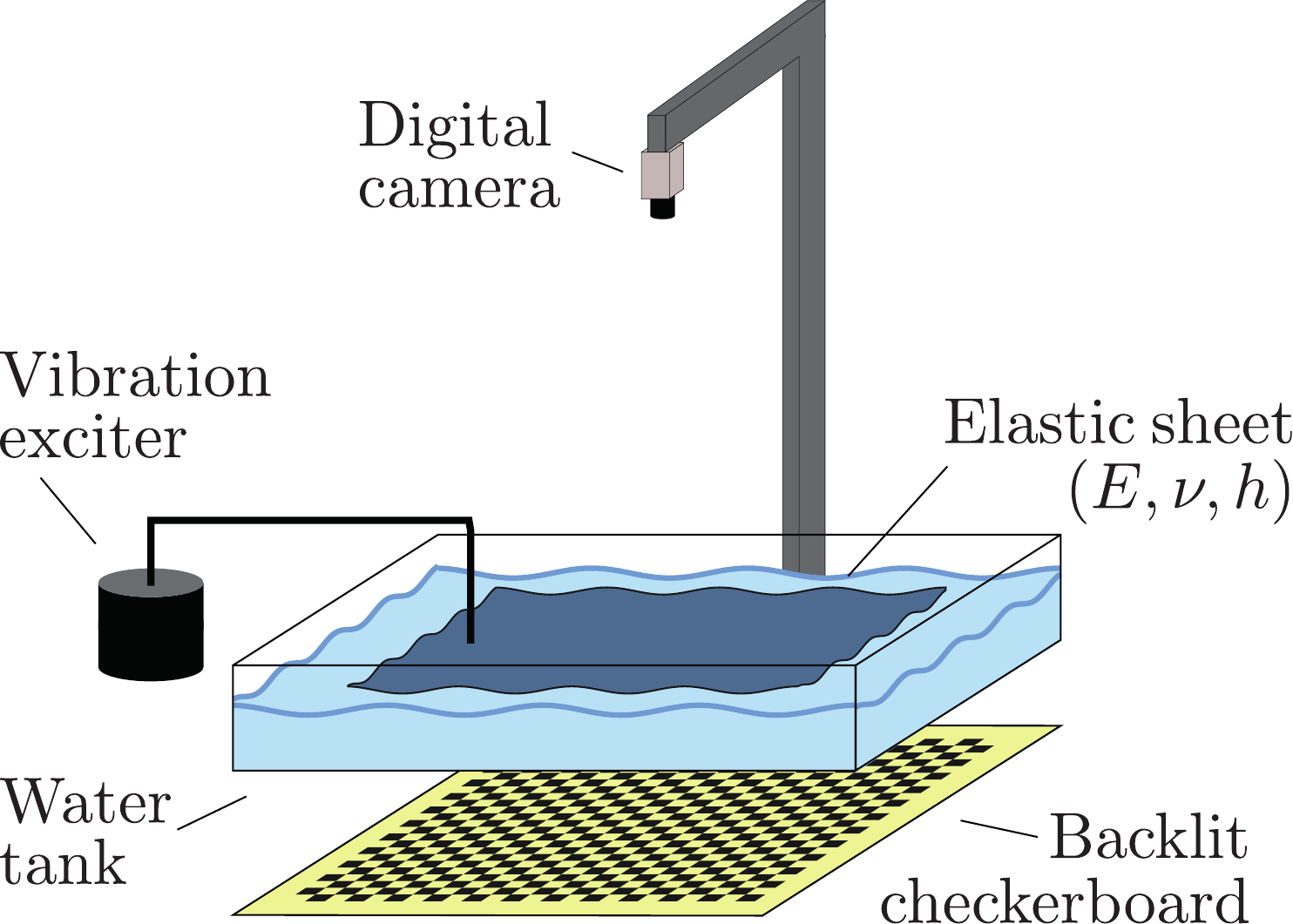}
    \caption{Sketch of the experimental set-up}
    \label{fig:set-up}
\end{figure}

A water tank of size 103 cm $\times$ 60 cm is filled with $h=17~$cm of water, to ensure that the surface waves propagate in the deep water regime. A clean silicone membrane with Young's modulus $E\simeq 1.6~$MPa and thickness $e = 400\ \mu$m is then placed at the water surface with free boundary conditions on the edges to avoid any extensional forces. To excite surface waves, we use a shaker (Bruel\ \& Kjaerr 4808), whose frequency $f_0$ can be varied from 4 to 160 Hz (figure \ref{fig:set-up}). To measure the surface displacement $\eta(x,y,t)$, we use the Free-Surface Synthetic Schlieren technique~\cite{Moisy_2009} and its later improvement of the Fast Checkerboard Demodulation~\cite{Wildeman2018}.
To do so, a checkerboard of 2 mm squares is placed under the tank and illuminated from below by a LED panel. 
For any surface gradient $\nabla \eta$, the checkerboard is displaced by a distance proportional to the gradient provided that the amplitude of the waves remains sufficiently small. This measurement technique has been calibrated previously and provides surface measurements with a typical accuracy of 5 $\mu$m under current experimental conditions. Using a CCD camera (Basler ac2048), we record movies of 150 square images with a sensor of 2048 $\times$ 2048 pixels, corresponding to a field of view 37.7 x 37.7 cm, which is larger than the largest imposed wavelength of $\lambda \sim 10$~cm at 4 Hz. 
For each forcing frequency, we carefully select a sampling frequency between 40 and 80 Hz to ensure stroboscopic imaging. With a recording typically lasting 2 to 4 seconds, we eventually acquire at least 50 different phases of the motion. The recordings are then processed using the Fast Checkerboard Demodulation algorithm (FCD)\cite{Wildeman2018}, which converts the checkerboard distortion in each image of the recording into an elevation field $\eta(x,y,t)$.  This algorithm is now routinely used to measure liquid interface waves\cite{Domino_2018, Domino_2020} and elastic waves \cite{Chantelot2020}. To extract the wave field specifically at the forcing frequency $f_0$, we compute the complex temporal Fourier transform $\hat{\eta}(x,y,f_0)$. The field $\mathcal{R}( \hat{\eta}(x,y,f_0) e^{i\phi})$ corresponds to the wave field at the phase $\phi$ of the oscillation at the frequency $f_0$.

\section{Dispersion relation for uniform sheets}

\begin{figure*}[t]
    \centering
\includegraphics[width=1\linewidth]{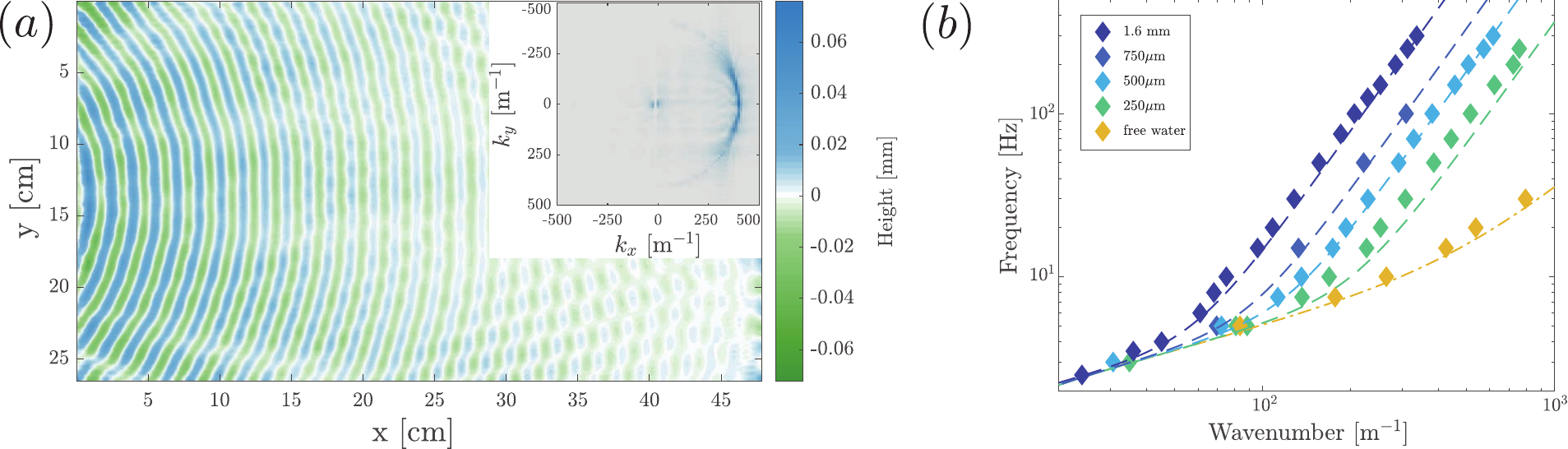}
    \caption{(a) Demodulated wave field height at one phase of the oscillation, $\mathcal{R}(\hat \eta(x,y,f_0))$ generated by a point source oscillating at $f_0=100$~Hz on a membrane of thickness $e=500~\mu$m. Inset: Spatial (2D) Fourier transform of the demodulated field $\mathcal{R}(\hat \eta(x,y,f_0))$. (b) Measured dispersion relation for uniform sheets with $E=1.6~$MPa, $\nu=1/2$ and thicknesses $e=250~\mu$m, $500~\mu$m, $750~\mu$m and $1.6~$mm respectively (see legend). The yellow diamonds correspond to a free-water surface. The dashed lines are the prediction from equation \ref{eq:disprel} without any adjustable parameters.}
    \label{fig:raw_waves}
\end{figure*}

For a uniform sheet, we first measured the dispersion relation for five different thicknesses following Domino {\textit et al.}~\cite{Domino2018}. Figure \ref{fig:raw_waves}(a) shows, in a color-coded plot, an example of the demodulated wave field height at one phase of the oscillation, $\mathcal{R}(\hat \eta(x,y,f_0))$. The source located on the left generates circular waves that propagate through the tank. From the complex field $\hat \eta(x,y,f_0)$, we compute the spatial Fourier transform, as shown in the inset of figure~\ref{fig:raw_waves}(a). The magnitude of each Fourier component is color-coded in blue. We observe that the energy is located on an arc length of circle in the Fourier space, showing that we measured circular propagating waves. From this map, we extract the wavenumber corresponding to the peak of energy, 
and we plot in figure~\ref{fig:raw_waves}(b) the resulting wavenumbers for various forcing frequencies, which altogether construct the dispersion relation of the waves. The colors correspond to 4 different membranes. Yellow symbols represent the dispersion relation of free-water, in which the restoring forces are only gravity and surface tension. We observe that the membrane dispersion relations lie above the free-water case. At higher frequencies, we recover the 5/2 exponent, corresponding to pure flexural waves. The theoretical dispersion relation for a floating membrane has been derived analytically for a perfect fluid, an infinitely thin elastic sheet with no inertia, and small vertical displacements~\cite{Schulkes1987,Davys1985}. It reads:
\begin{equation}
    \omega^2 = gk+\frac{T}{\rho}k^3 + \frac{D}{\rho}k^5,
    \label{eq:disprel}
\end{equation}
where $\omega$ is the angular frequency, $g=9.81~$m.s$^{-2}$ the Earth's gravity, $\rho$ is the fluid density, $k = \frac{2\pi}{\lambda}$ is the wavenumber, $T$ is the membrane tension which is set by the surface tension of water for a free-floating membrane, and $D$ is the flexural modulus. This modulus is obtained from F\"oppl-von Karman equation and is given by:
\begin{equation}
    D = \frac{Ee^3}{12\left(1-\nu^2\right)},
\end{equation} 
where $E$ is the Young's modulus and $\nu$ is the Poisson ratio.  The theoretical prediction is superimposed on the experimental measurements shown in figure~\ref{fig:raw_waves} as colored dashed lines. The model shows quantitative agreement with the experimental data without any adjustable parameters. For all sheets, we observe two regimes. At low $k$, gravity dominates and $\omega^2=4*\pi^2*f^2 = g k$ whereas bending is the main restoring force at large $k$ where $\omega^2\sim D/\rho k^5$. We introduce 
\begin{equation}
l_D=\left(\frac{D}{\rho g} \right)^{1/4}
\end{equation}
as the gravito-elastic lengh of each sheet. The separation in between the two regimes occurs for $k=1/l_D$. Note that in practice for a sheet floating at the water surface, the tension term is equal to the surface tension of the air-water interface. The tension term is then usually much smaller than the gravity and bending term.

\section{Local k-extraction}

Extracting local properties of the medium using variations of the dispersion relation has been proposed in various contexts. In particular, elastography imaging has faced the  problem of elasticity extraction from an elastic wave field since the 90\'s. In this context, different methods have emerged: a wave front  time-of-flight approach mainly used in ultrasound elastography \cite{catheline1999a, catheline1999b, sandrin1999} or correlation-based method from diffuse field (used both in optical elastography \cite{schmitt1998, marmin2020, marmin2021} and seismology \cite{campillo2003}). Another proposed approach relates to an inverse problem of the Helmholtz equation used in magnetic resonance elastography \cite{muthupillai1995}, aiming at retrieving the local variations of the medium's elasticity \cite{oliphant2001}.

To extract the local flexural modulus of the membrane, we aim at estimating locally the wave dispersion relation. In general, for surface waves, in the presence of varying spatial properties for the wave propagation such as the water depth or the flexural modulus $D$, there is no exact wave equation, due to the non-locality of the pressure term in incompressible flows. However, we can still experimentally compute a local wavenumber at each point and assume an equivalent dispersion relation to estimate the flexural modulus $D$. \\

    To illustrate our method, we consider a uniform thin plate, following the dispersion relation given by Eq.~\ref{eq:disprel}. The temporal Fourier transform $\hat \eta(x,y,\omega)$ of the surface height is a solution of a Helmholtz equation:
    \begin{equation}
    \left(\Delta+k^2(\omega)\right)\hat{\eta} =0,
    \end{equation}
    where $k(\omega)$ is given by the dispersion relation of hydro-elastic waves.
We then decompose the spatial field $\hat \eta$ into a base of functions of the Helmholtz equation, namely the Bessel functions $J_n$. Graf's theorem states that the solutions of the Helmholtz equations can be expanded in Bessel function series centered around any point $\bm{r}$ in space. Using a polar coordinate system $(R,\theta)$ centered around the point $\bf r$ of interest, we deduce that $\hat{\eta}$ can be written as:
\begin{equation}
    \hat{\eta}(R,\theta,\omega) = \displaystyle{\sum_{n=-\infty}^{+\infty}}a_nJ_n(k R){\rm e}^{in\theta},
\end{equation}

\begin{figure}
    \centering
\includegraphics[width=1\columnwidth]{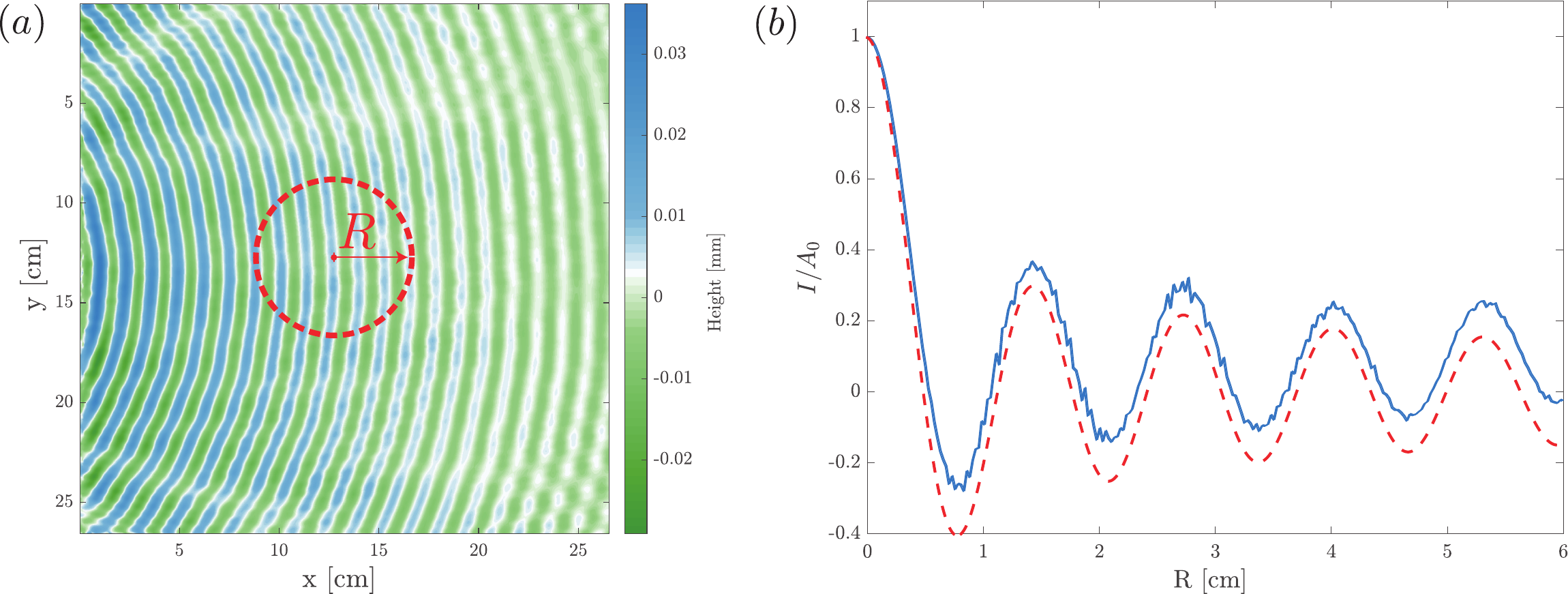}
    \caption{Principle of extraction of $I(x,y,R)$ for a membrane with thickness $e=500~\mu$m at $f_0=150$~Hz.(a) The integral $I(x,y,R)$ is computed by integrating the wavefield along a circle of radius $R$ centered at location $(x,y)$. (b) Experimental measurement of $I(R)$ (blue solid line) for the wavefield showed in fig.(a). The profile matches a Bessel function (red dashed line), which indicates the local value of $k$.}
    \label{fig:champ_extrait}
\end{figure}

where $a_n$ are complex number coefficients. To extract the local wave number around $\bf{r}$, we notice that the integral $I$ of $\hat{\eta}$ on a circle of radius $R$
\begin{equation}
I(R) = \dfrac{1}{2\pi}\displaystyle{\oint_R}\hat{\eta}(R,\theta, \omega)d\theta,
\end{equation}
is related to the local wave number $k$ of the height field. Indeed, using the Bessel decomposition centered around $\bf{r}$, we have
\begin{equation}
I(R) = \frac{1}{2\pi} a_0 J_0(kR),
\label{eq:integral}
\end{equation}
where $R$ is the distance to the point $\bf{r}$. The method is illustrated in figure~\ref{fig:champ_extrait} with the height field $\eta$ obtained with a membrane of thickness $e = 500~\mu$m excited with a wave of frequency $f_0 = 150$~Hz. The dashed red circle used for computing the integral $I$ for a radius $R=3$cm is superimposed. Figure~\ref{fig:champ_extrait} shows the normalized integral $I/A_0$, where $A_0 = a_0/(2\pi)$ as a function of $R$. We observe that $I/A_0$ is indeed a Bessel function $J_0$, and a fit gives the value of the wavenumber $k$ corresponding to the dispersion relation of hydro-elastic waves for $f_0$. In practice, we fit only the local profile of $I/A_0$ near $R=0$ using a Taylor expansion of $I$ around 0:
\begin{equation}
I(R) = 1 - J_0''(0) k^2 r^2/2,
\label{eq:integral_expansion}
\end{equation}
to extract the value of $k$. We optimize the number of points used in the parabolic fit. Increasing the number of points enhances the robustness against noise but reduces spatial resolution. In practice, the parabolic fit is performed on the first 6 points, to ensure an optimal balance between robustness and spatial resolution. For the case of figure~\ref{fig:champ_extrait}, the spatial resolution on the wavenumber extraction is typically one millimeter. Note that this value is smaller than Abbe diffraction limit, leading somehow to super-resolution. This is not surprising in this context \cite{Zemzemi2020}, as the resolution is proportional to the pixel size of the reconstructed wavefield. This value is independent from the wavelength $\lambda=2*pi/k$ which is always larger than $8~$mm in our experiments. The method extends to heterogeneous elastic sheets, with only locally homogeneous mechanical properties. Indeed, considering homogeneous subdomains of arbitrary shapes separated by sharp interfaces, the Helmholtz equation is valid in each subdomain, and the contour integral method works within each subdomain. In the case of continuous variations of the flexural modulus, however, the Helmholtz equation is not strictly valid, and the method will only provide an estimate of the local flexural modulus, with corrections depending on the flexural modulus gradient $\nabla D/(D~k)$.


\begin{figure}[t!]
    \centering
    \includegraphics[width=1\columnwidth]{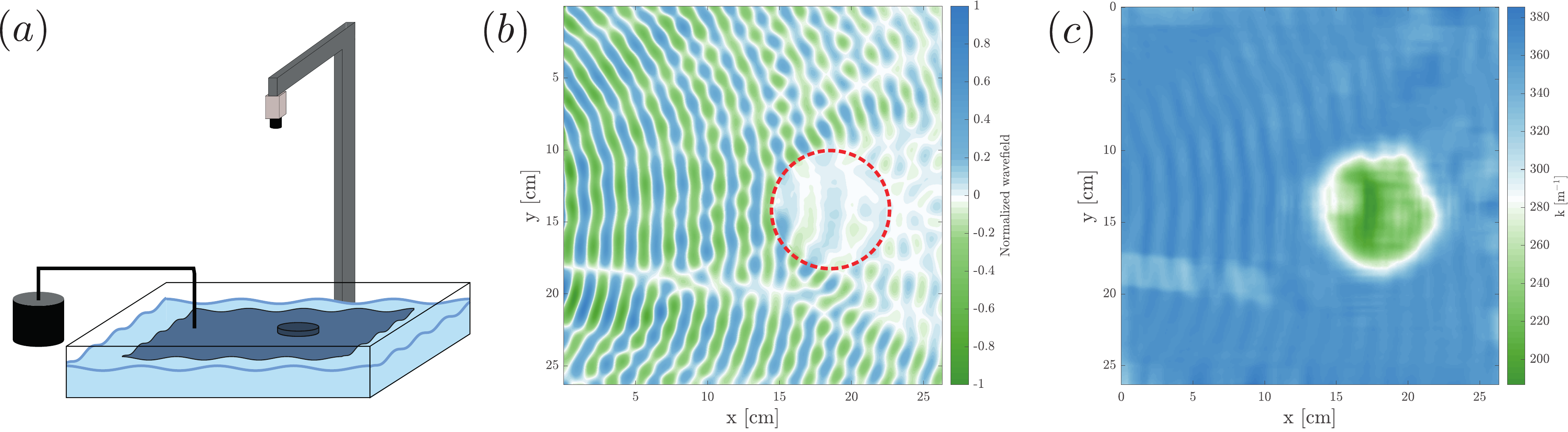}
    \caption{(a) Sketch of the experimental set-up: a circular patch is added on top of a uniform membrane. (b) Measured wavefield for a membrane with thickness $e=500~\mu$m and a patch of diameter $R=3.6~$cm and additional thickness $e_{disk}=1.6~$mm at forcing frequency $f_0=~75$Hz. (c) Map of extracted values of local wavenumber $k$.}
    \label{fig:circular_patch}
\end{figure}

\section{Measuring the thickness of local patches}

We now consider a sheet with spatial variations of the flexural modulus $D$, which can originate from the variation of the Young's modulus, the Poisson ratio, or the local thickness. The expression of $I(R)$ given by eq.~\ref{eq:integral} is no longer valid everywhere, as $\hat \eta$ is not a solution of the Helmholtz equation. However, near $R=0$, we assume that $I(R)$ still varies quadratically with the expression given by eq.~\ref{eq:integral_expansion}, with a wavenumber given by the equivalent dispersion relation associated with the mechanical properties of the membrane in $R=0$.

\begin{figure*}[t!]
    \centering
    \includegraphics[width=.8\columnwidth]{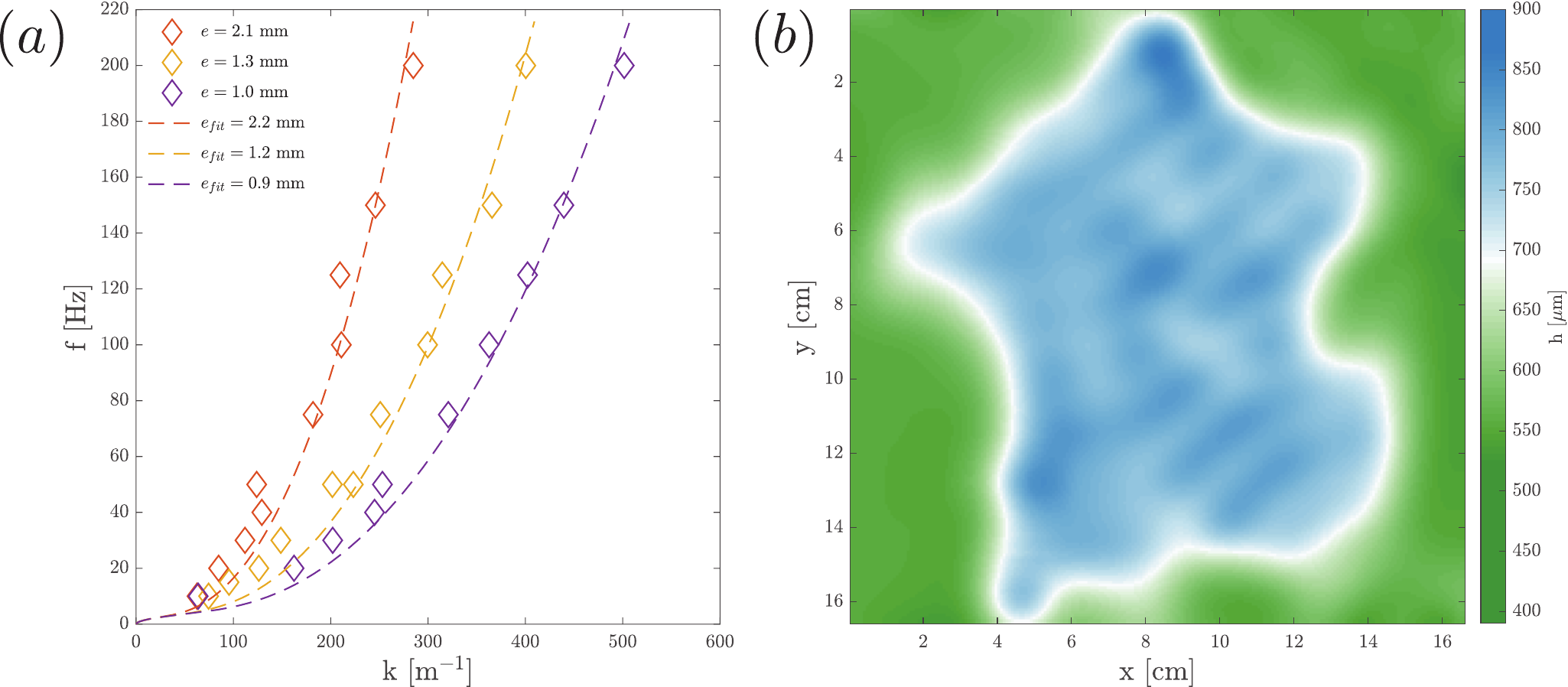}
    \caption{(a) Measurement of the local wavenumber for 3 different circular patches with total thickness $e_{tot}=1.0, 1.3$ and $2.1~$mm for frequencies varying from $f_0=10~$Hz to $200~$Hz. The dashed line is the fitted dispersion relation that allows the extraction of the local thickness inside the disk. (b) Tomographic measurement of the thickness. Thin part $e=400~\mu$m, Thick part $e=800~\mu$m. Measurement made at 160 Hz, with 2 point sources.}
    \label{fig:extracted_thickness}
\end{figure*}

To test the method, we only consider (for practical reasons) local variations of sheet thickness, but the method could be equally applied to variations of Young's modulus or Poisson ratio. We first test the method by gluing circular patches of the same material on the uniform membrane. Doing so, we vary spatially the flexural modulus $D$ through the local thickness.
The experimental procedure is sketched in figure~\ref{fig:circular_patch}a). 
A patch of thickness $e = 1.6$~mm and diameter $d = 7.2$~cm is glued onto the surface, at the location indicated by the dashed red circle in figure~\ref{fig:circular_patch}b. 
As a consequence, the space is divided into two regions, of thicknesses $e_1 = 500~\mu$m and $e_2 = 2100~\mu$m, corresponding to different flexural moduli $D_1$ and $D_2$, with a ratio $D_1/D_2 = (e_1/e_2)^3$. 
Qualitatively, we indeed observe a drastic increase of the wavelength inside the patch due to the variation in elastic properties, which modifies the dispersion relation. We then apply our method to each point in space. For $k \ell_D \gg 1$, the dispersion relation is dominated by the bending term, and writes:
\begin{equation}
\omega^2 = \frac{D}{\rho} k^5,
\end{equation}
where $D$ is the local flexural modulus. We compute the local membrane thickness variations $e$ compared to a reference state as :
\begin{equation}
\frac{e_1}{e_2} = \left(\frac{k_1}{k_2} \right )^{-5/3}.
\label{eq:thickness}
\end{equation}
The map of membrane thickness $e$ computed from the local extraction of the wavenumber and eq.~\ref{eq:thickness} is shown in figure~\ref{fig:circular_patch}. We recover the shape and the size of the circular patch glued on the surface. We analyse the reconstructed shape using a thresholding algorithm. We first determine the barycenter of the shape and then extract the average value of the radius in bins of $15^{\circ}$. We obtain an average value $R_{exp}=3.92$~mm with a standard deviation $\sigma=.17~$mm that shall be compared with the prescribed value $R=3.6$~mm. Our measurement is slightly overestimating $R$ but falls within a $2\sigma$ confidence interval. In order to verify the quantitative estimate of $D$, we measure the wavenumber inside the circular patch as a function of the forcing frequency, for three circular patches of thickness $e= 500, 800, 1600~\mu$m. The resulting dispersion relations are shown in figure~\ref{fig:extracted_thickness}(a) with colored diamond symbols. We then fit the dispersion relation using the theoretical prediction for hydro-elastic waves [eq. \ref{eq:disprel}] and extract the thickness using the expression of the flexural modulus, considering that the other membrane parameters (Young modulus and Poisson ratio) remain constant. We find a quantitative agreement with less than 10\% of errors on the membrane thickness.

We eventually test the method on a more complex shape, that does not show any axis of symmetry. We glue a patch of thickness $e = 400~\mu$m in the shape of metropolitan France on top of a uniform membrane of thickness $e=400~\mu$m, hence doubling locally the sheet thickness. Surface waves are excited in the bending regime at the frequency $f_0 = 160$~Hz from 2 different point sources located on each side of the patch. We then apply the wavenumber extraction method on the surface height measured in the region of the patch. The resulting extracted thickness map is shown in figure~\ref{fig:extracted_thickness}(b). We recover the initial shape with a sub-wavelength accuracy and find a quantitative agreement with the patch thickness. 

    
\section{Conclusion}

We present a method to extract the local wavenumber of a wave field, and we apply this technique to a spatially varying elastic sheet. We show that from spatio temporal measurement of the wave field, we can quantitatively recover the dispersion relation of surface waves. In the bending regime of wave propagation, we show that the local membrane flexural modulus $D$ can be extracted. For a homogeneous material with a constant Young's modulus $E$, we can thus infer the local thickness $e$. We implemented the technique on two model cases, circular patches of different thicknesses, and a complex shape patch. In both cases, we show that the local thickness can be quantitatively measured, with a subwavelength accuracy.

\acknowledgments{A.E. would like to thank F. Lemoult for stimulating discussions. This work has benefited from the financial support of Mairie de Paris through Emergence(s) grant 2021-DAE-100 245973, and from the Agence Nationale de la Recherche through grant MSIM ANR-23-CE01-0020-02.}

\end{document}